\documentclass[10pt,english,prl,showpacs,superscriptaddress]{revtex4}
\usepackage[latin9]{inputenc}
\usepackage{textcomp}
\usepackage{amsmath}
\usepackage{amssymb}
\usepackage{graphicx}

\makeatletter
\@ifundefined{textcolor}{}
{%
 \definecolor{BLACK}{gray}{0}
 \definecolor{WHITE}{gray}{1}
 \definecolor{RED}{rgb}{1,0,0}
 \definecolor{GREEN}{rgb}{0,1,0}
 \definecolor{BLUE}{rgb}{0,0,1}
 \definecolor{CYAN}{cmyk}{1,0,0,0}
 \definecolor{MAGENTA}{cmyk}{0,1,0,0}
 \definecolor{YELLOW}{cmyk}{0,0,1,0}
}

\usepackage{epstopdf}
\usepackage{color}
\makeatother

\usepackage{babel}
\begin{document}

%%%% Article title to be placed here
\title{Extended and localized Hopf-Turing mixed-mode in non-instantaneous Kerr cavities}

\author{M. Ouali,$^{1}$ S. Coulibaly,$^{1}$ M. Taki,$^{1}$ and M. Tlidi,$^{2}$}

%%%%%%%%% Insert author address here
\address{$^{1}$ Univ. Lille, CNRS, UMR 8523, PhLAM, Physique des Lasers Atomes et Mol\'ecules, F-59000 Lille, France
\\
$^{2}$Facult\'e des Sciences, Universit\'e Libre de Bruxelles (ULB), Code Postal
231, Campus Plaine, Bruxelles B-1050, Belgium}

\begin{abstract}
We investigate the spatio-temporal dynamics of a ring cavity filled with a non-instantaneous Kerr medium and driven by a coherent injected  beam. We show the existence of a stable mixed-mode solution that can be either extended or localized in space. The mixed-mode solutions are obtained in a regime where Turing instability (often called modulational instability) interacts with self-pulsing phenomenon (Andronov-Hopf bifurcation). We numerically describe the transition from stationary inhomogeneous solutions to a branch of mixed-mode solutions. We characterize this transition by constructing the bifurcation diagram associated with these solutions. Finally, we show stable localized mixed-mode solutions, which consist of time-periodic oscillations that are localized in space.
\end{abstract}

\maketitle
\section{Introduction}
Transition from spatially uniform state to a self-organized or ordered structures is a universal feature in far from equilibrium systems and has been observed in many natural systems  (see overviews on this issue  \cite{Rev12,Rev13,Leblond-Mihalache,Tlidi-PTRA,Tlidi_Clerc_sringer}).  This transition is triggered by a few number of modes often called Turing modes that leads to the formation of spatially stationary structures that can be either periodic or localized in space.  On the order hand,  a transition to a time oscillation or a self-pulsing state through the Andronov-Hopf (termed Hopf in the following) bifurcation has also been  observed.  These transitions are responsible for the symmetry breaking  in space (Turing) and in time (Hopf). The interaction between Turing and Hopf bifurcations  concerns almost all fields of nonlinear science such as biology, chemistry, physics, fluid mechanics, and optics. When both instabilities are close one to another, the space-time dynamics  of various spatially extended systems may be significantly impacted. In particular, it has been shown that in this regime either a bistable behavior between Turing patterns and self-pulsing states  or front waves between Hopf and Turing-type domains may occur \cite{Bistab_HOpf_TURING2,Bistab_HOpf_TURING3,Bistab_HOpf_TURING4,Bistab_HOpf_TURING5,Barcella,Martine_JOPB_2004}. Moreover, mixed-mode solutions resulting from the interplay between Andronov-Hopf and Turing modes may dominate the spatio-temporal dynamics  in  optical frequency conversion systems \cite{Tlidi_haelterman97,Tlidi_haelterman98}.   This interaction can also  lead to a spatio-temporal chaos  \cite{Dewel2}. 

In this paper, we consider a ring cavity filled with a non instantaneous and a nonlocal Kerr medium, and driven by a coherent injected beam. We show  that  the threshold associated with Turing and Hopf bifurcations can occur arbitrarily close one to another leading to codimension-two point where both bifurcations coincide. In the monostable regime, we show that the Turing pattern becomes unstable and leads to the formation of a Turing-Hopf mixed-mode. More importantly, we show that localized mixed-mode solutions can be stabilized when the system exhibits a coexistence between an homogeneous steady state and an extended mixed-mode solution.   
\section{Model equations and a linear stability analysis}
The slowly varying envelope of the electric field $E=E(x,t)$ circulating inside the cavity is described by $\partial E/\partial z=inE+i\partial^2 E/\partial x^{2}$ coupled with an additional equation for the  material photoexcitation $n=n(x,t)$: $\gamma^{-1}\partial n/\partial z=|E|^2+d\partial^2 n/\partial x^{2}$ with $z$ is
the longitudinal coordinate, and  the parameter $d$ is proportional to the ratio between the diffusion length  and diffraction coefficients.  This parameter describes the degree of nonlocality. Indeed, if $d = 0$, the above propagation equations have a local nonlinear response. The strong nonlocal response is characterized by a large $d$. The parameter $ \gamma$ is proportional to the ratio between the characteristics decay times associated with the electric field and the refractive index $n$, respectively.  At each round trip, the light inside the Kerr media is coherently superimposed with the input
beam. This can be described by the cavity boundary conditions $E(z=0,x)=rS+\tau E(L,x)\exp{(i\phi_e)}$ and  $E(z=0,x)=rn(L,x)\exp{(i\phi_n)}$, with $\phi_{e,n}$ are linear phase shifts,  $S$ is the ampliude of the injected field, and $L$ denotes the  cavity length. The letters "$r$" and "$\tau$" indicate the transmission and the reflexion coefficients. We adopt the well know mean field approach proposed by Lugiato and Lefever \cite{LL}. A detailled calculations of the derivation of the mean field model can be found in the appendix of the paper \cite{Kockaert_2006}. This approach  is valid under the
following approximations:~the cavity possesses a high Fresnel number and
we assume that the cavity length is much shorter than the diffraction, diffusion and the nonlinearity
spatial scales. A single longitudinal mode operation is also assumed. The mean field approach  applied to our system leads to the adimensional coupled partial differential equations
\begin{eqnarray}
\frac{\partial E}{\partial t} &=& S - (1+i\delta)E + inE + i\frac{\partial^2 E}{\partial x^{2}},  \label{eq:dEdt} \\
 \gamma \frac{\partial n}{\partial t} &= & -n+|E|^2  + d \frac{\partial^2 n}{\partial x^{2}},   \label{eq:dndt}
\end{eqnarray} 
where $\delta$ is the frequency detuning between the injected light and the cavity resonance. $S$ is the injected field amplitude considered as being real without loss of generality.  Diffraction of the electric field in the transverse $x$ coordinate is modeled by $i\partial^2 E/\partial x^{2}$. The density of material photoexcitation is denoted by $n$ that diffuses according to $\partial^2 n/\partial x^{2}$. In the limit of $ \delta=0$, the model 
has been derived for a Kerr medium but in a single feedback mirror configuration \cite{Firth90}.  For a ring cavities, and in the particular limit where $d=0$ and $ \gamma=0$, one obtains from Eq. (\ref{eq:dndt}) $n=|E|^2$. By replacing $n$ by $|E|^2$ in Eqs. (\ref{eq:dEdt}), we recover the well known Lugiato-Lefever (LL) equation \cite{LL}.  The model Eqs. (\ref{eq:dEdt}) and \ref{eq:dndt}) may be viewed as an extension of LL equation to include the non instantaneous response of the medium and the nonlocal effects.

Homogeneous steady state (HSS) solutions of Eqs.~(\ref{eq:dEdt}) and (\ref{eq:dndt}) are found by setting the time and space derivatives equal to zero: $n_s=I_s$ and $S^2=I_s[1+(\delta-I_s)^2]$ with $I_s=|E_s|^2$.  The characteristic $I_s$ as a function of the input intensity $S^2$ is  monstable when 
$\delta<\sqrt{3}$. The HSS's undergo a bistable behavior when  $\delta>\sqrt{3}$.  The linear stability analysis of the HSS with respect to a finite wavelength perturbation of the form  $\exp{(\lambda t + ikx)}$ leads to the third-order polynomial characteristic equation:
$a_0 +a_1 \lambda+a_2 \lambda^2+a_3 \lambda^3=0$, with $a_0=1-I_s^2+(I_s-Q)^2+dk^2(1+Q^2)$, $a_1=2(1+dk^2)+\gamma(1+Q^2)$, $a_2=1+2 \gamma +dk^2$ and $a_3=\gamma $, with $Q=\delta-I_s+k^2$. The Turing type of bifurcation occurs when  $\lambda=0$ and $\partial \lambda /\partial k^2=0$. The threshold associated with this instability and the critical wavelength at the Turing bifurcation are obtained from the expression $a_0=0$. This instability is characterized by an intrinsic wavelength which is determined by the dynamical parameters and not by the external geometrical constraints such as boundary conditions. 

A Hopf bifurcation occurs if a pair of complex-conjugate roots has a vanishing real part and  a nonzero imaginary part. In the limit $\gamma=1$ and $\delta <2$, only one Hopf bifurcation point is possible.  The associated critical intensity is explicitly given by  
$I_H=2(2-\delta d)(d+\sqrt{d^2+1})$. At this bifurcation point the real part vanishes at nonzero wavenumber $k_H^2=-\delta+(2-\delta d)/\sqrt{d^2+1}$. Note that the HSS's may undergo a homogeneous Hopf bifurcation  where the real part of the eigenvalues vanishes for  $k=0$.

We are interested in the regime where Turing and Hopf instabilities are close one to another. For this purpose, we fix $\gamma=1$ and $\delta=1$, and we vary the coefficient $d$ and the input field intensity.  An important and interesting feature is that the parameter $d$ controls the relative position of the thresholds associated with both Turing and Hopf instabilities. Typical marginal stability curves are shown in Fig. 1.  By increasing the value of the parameter $d$, the first bifurcation is of Turing type,  and the Hopf bifurcation appears as secondary instability as shown in Fig. 1(a). There exists a critical value of the parameter $d$ for which both bifurcations coincide, i.e., $I_H=I_T=I_c$  as shown in Fig. 1(b). At this co-dimensional two point, one of the real roots of the characteristic equation  vanishes $\lambda=0$ with a non zero finite intrinsic wave length $2\pi/k_c$ and two other roots are purely imaginary   $\lambda=\pm i \omega $. Note that the critical mode $k_c$ is also$d$-dependent. When further increasing $d$,  the first instability is the Hopf bifurcation followed by the Turing instability as shown in Fig. 1(c).
\begin{figure}[h]
\centerline{\includegraphics[width=12cm]{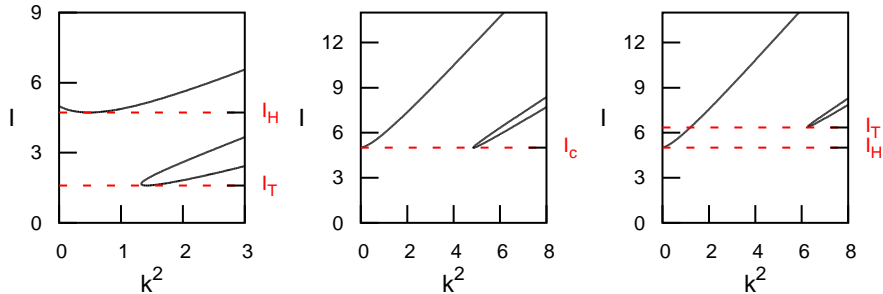}}
\caption{Marginal stability curves. The parameters are $\gamma=1$, $\delta=1$ with $I_H$ and $I_T$ are Hopf and Turing instability thresholds, respectively. (a) $d=0.4$, above the upper curve HSS is Hopf unstable. Inside the lower curve, HSS is Turing unstable. (b) $d=0.81$, note the co-dimension two instability where Turing and Hopf thresholds coincide. (c) $d=0.85$, the situation is inverted with respect to (a).}
\label{MS}
\end{figure}
\begin{figure}[htbp]
\centerline{\includegraphics[width=6cm]{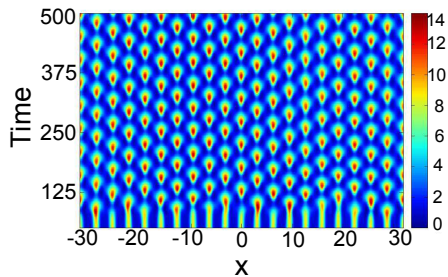}}
\caption{Space-time map showing the formation of a Turing-Hopf mixed-mode solution. The parameters are $\gamma=1$, $\delta=1$, $S=5.28$ and $d=0.4$.}\label{MM}
\end{figure}
\begin{figure}[htbp]
\centerline{\includegraphics[width=6cm]{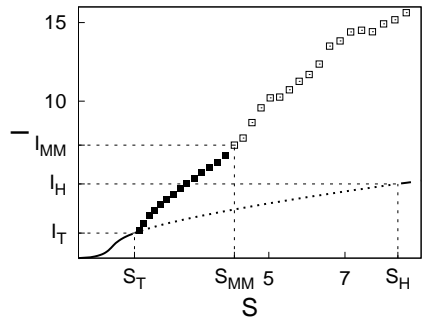}}
\caption{Bifurcation diagram showing the evolution of the intracavity field intensity as a function of the input field amplitude.  Stable (unstable) homogeneous steady states are indicated by solid (dotted) line. Filled squares denote the maximum intensity of the stationary periodic structures. Empty squares indicate the maximum intensity associated with mixed-mode solutions. Parameters are $\gamma=1$, $\delta=1$, and $d=0.4$.}\label{DiaBifmon}
\end{figure}
\begin{figure}[htbp]
\centerline{\includegraphics[width=8cm]{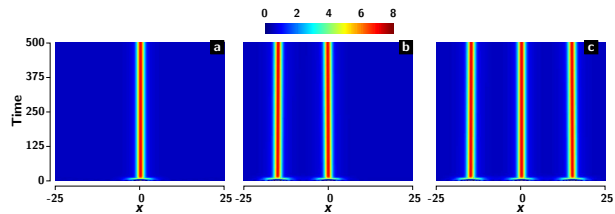}}
\caption{Space-time maps displaying  the formation of multi-peaks stationary localized structures. (a) one, (b) two and (c) three peaks  obtained by numerical simulations of Eqs. (1) and (2) for the parameters  $\gamma=1.3$, $\delta=3$, $S=1.9$, and $d=1.45$.}\label{LS}
\end{figure}
\begin{figure}[htbp]
\centerline{\includegraphics[width=10cm]{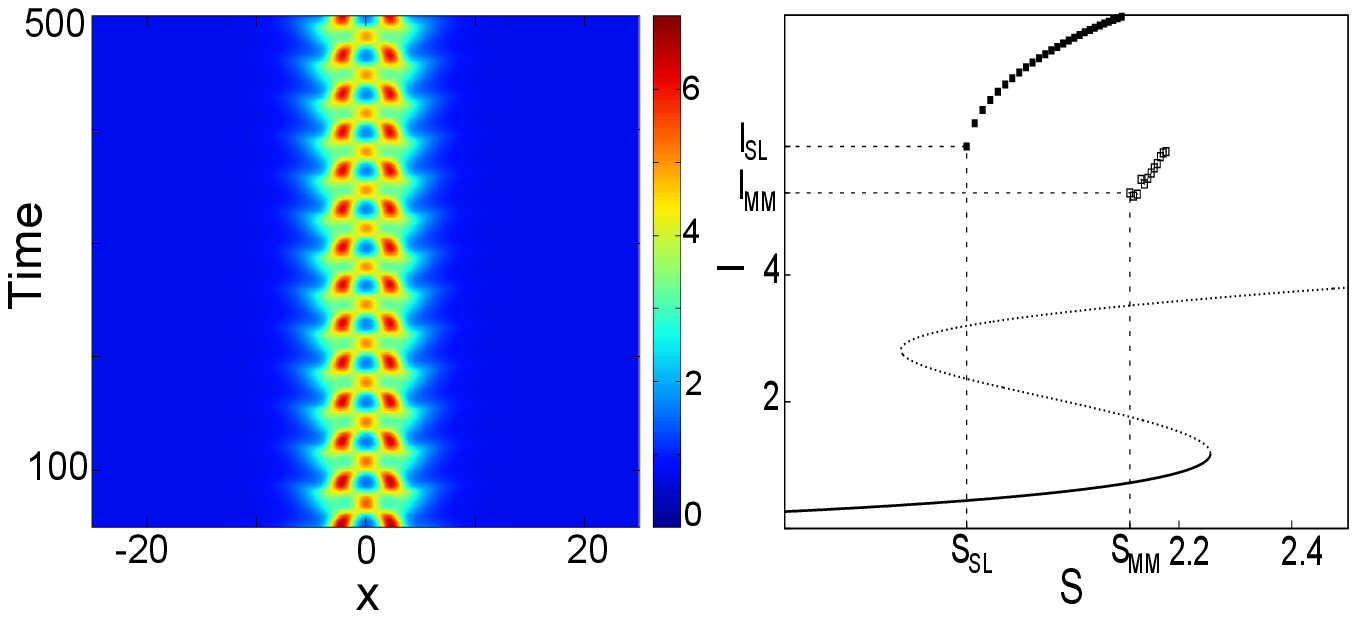}}
\caption{(a) Space-time map showing the generation of a localized mixed-mode solution. Parameters are $\gamma=1.3$, $\delta=3$, $S=2.11$, and $d=1.45$. (b) Bifurcation diagram showing the evolution of the intracavity field intensity as a function of the input field amplitude.  Stable (unstable) homogeneous steady states are indicated by solid (dotted) line. Filled squares denote the maximum intensity of the stationary localized  structures. Empty squares indicate the maximum intensity associated with  localized mixed-mode solutions. Same parameters as in (a) with varying $S$.}\label{LMM}
\end{figure}
\section{Extended and localized mixed-mode}
In what follows we focus on numerical investigations of the model Eqs. (1) and (2)  by using an adaptive step-size  Bulirsch-Stoer method \cite{Hairer_1993}. We choose parameters values such that the homogeneous steady state is destabilized first by  Turing instability.   From this bifurcation point, a branch of stationary spatially periodic solutions appears supercitically with a well defined wavelength. However, when increasing further the injected field intensity a stable mixed-mode solution is spontaneously generated in the system. These solutions correspond to oscillations both in time and in space. A typical  example of such a behavior is plotted in Fig. 2. To characterize the transition from stationary periodic patterns to a branch of mixed-mode Turing-Hopf structures, we plot in Fig. 3 the maxima of both Turing structures and mixed-mode solutions together with the homogeneous steady state. As we increase the amplitude of the injected field, the HSS becomes unstable, and a spatially periodic structure emerges from the Turing bifurcation point $S=S_T$. These structures are stable in the parameter range $S_T<S<S_{MM}$.  For $S>S_{MM}$, the Turing structure becomes unstable and bifurcates to a stable mixed-mode solution.

The extended solutions either Turing structures or mixed-mode solutions are obtained in the supercritical regime. Besides these extended structures there exists another type of solutions which are aperiodic but localized in space (LS's). The latter are found in the  subcritical regime associated with the Turing instability. It is well known that the emergence of these solutions does not necessarily requires a bistable regime \cite{Tlidi_94}. The   prerequisite condition for the generation of LS's is the coexistence between a single HSS and the spatially periodic solutions. This solutions have been predicted theoretically  \cite{Scorggie_csf94} and experimentally evidenced in an instantaneous and local optical Kerr medium \cite{Majid_NJP}. Notice that temporal localized solutions known also as temporal solitons have also been experimentally observed in a dispersive Kerr cavity in \cite{Leo}. Here we show that in the case of a non-instantaneous and a nonlocal Kerr medium, modeled by the Eqs. (1) and (2), support localized structures. An example of such a behavior  is shown in Fig. 4. We show in this figure only localized structures with one, two and three peaks. The number of localized peaks and their spatial distributions are determined by the initial conditions \cite{Scorggie_csf94}. There exists an infinite number of spatially localized solutions if the size of the system is infinite.  When increasing the injected field intensity, localized structures exhibit a pulsing  phenomenon, i.e., time oscillations in a wide range of parameters leading to the formation of Hopf-Turing mixed-mode solutions.  A typical example of such a  behavior is shown in Fig. 5(a). The maxima of intensity associated with  localized structures and localized mixed-mode solutions are plotted together with the homogeneous steady states in Fig. 5(b). When increasing the input intensity,  a single peak stationary localized structure is formed in the range $S_{SL}<S<S_{MM}$. For $S>S_{MM}$, stable localized mixed-mode solutions are generated. Their maximum intensity  are shown in Fig. 5(b).  The width of these time dependent structures varies in the course of time.  
\section{Conclusions}
We have shown that extended mixed-mode solutions can be generated in the output of a driven ring cavity filled with a non-instantaneous and a nonlocal Kerr medium. Our investigations are focused on the regime where Turing and Hopf instabilities interact strongly. In the monostable regime, the mixed-mode solutions appear as a transition from spatially periodic structures.  We have drawn the bifurcation diagram showing their stability domain. In the bistable regime, we have shown occurrence of stationary localized structures that exhibit multistability in a finite range of parameters. Finally, we have established evidence of localized mixed-mode solutions that emerge from the branch of a single peak localized structures.  We have constructed a bifurcation diagram associated with these localized structures. 
\section*{Acknowledgments}
We are very grateful to the support from The "Centre National de la Recherche Scientifique (CNRS)", France. This research was partially supported by the Interuniversity Attraction Poles program of the Belgian Science Policy Office, under grant IAP P7-35 \textit{photonics@be}. The support of Ministry of Higher Education and Research as well as by the Agence Nationale de la Recherche through the LABEX CEMPI project (ANR-11-LABX-0007) is also acknowledge. M. T. received support from the Fonds National de la Recherche Scientifique (Belgium). 

\end{document}